\documentclass[11pt,a4paper]{article}

\usepackage{amsmath,amssymb}
\usepackage{axodraw4j}
\usepackage{epsfig,graphicx}
\usepackage{subfigure}
\usepackage{graphicx}
\usepackage{rotating}
\usepackage{cancel}
\usepackage{hyperref}
\usepackage{bm}
\usepackage{color}
\usepackage{cite}
\numberwithin{equation}{section}

\long\def\symbolfootnote[#1]#2{\begingroup%
\def\thefootnote{\fnsymbol{footnote}}\footnote[#1]{#2}\endgroup}

\def\beq{\begin{equation}}
\def\eeq{\end{equation}}

\begin{document}

\title{\vspace*{-1cm}\normalsize \hfill\hspace*{11cm}\mbox{IPPP/10/67; DCPT/10/134}\hfill\mbox{}\\[1.5cm]
\LARGE{\textbf{Spectroscopy as a test of Coulomb's law}}\\
\Large{\textbf{--}}\\
\Large{\textbf{A probe of the hidden sector}}}
\author{Joerg Jaeckel and Sabyasachi Roy
\
\\[2ex]
\
\small{\em Institute for Particle Physics Phenomenology,} \\
\small{\em Durham University, Durham DH1 3LE, United Kingdom}\\[1.5ex]
}
\date{}
\maketitle

\begin{abstract}

\noindent 
High precision spectroscopy can provide a sensitive tool to test Coulomb's law on atomic length scales.
This can then be used to constrain particles such as extra ``hidden'' photons or minicharged particles that are predicted in many extensions of the standard model, and which cause small deviations from Coulomb's law.
In this paper we use a variety of transitions in atomic hydrogen, hydrogenic ions, and exotic atoms
to probe Coulomb's law. 
This extends the region of pure Coulomb's law tests to larger masses. For hidden photons and minicharged particles
this region is already tested by other astrophysical and laboratory probes. However, future tests of true muonium and muonic atoms
are likely to probe new parameter space and therefore have good discovery potential for new physics.
Finally, we investigate whether the discrepancy between the theoretical calculation of the $2s_{1/2}^{F=1} - 2p_{3/2}^{F=2}$ transition in muonic hydrogen and its recent experimental measurement at PSI can be explained by the existence of a hidden photon.
This explanation is ruled out by measurements of the Lamb shift in ordinary hydrogen.
\end{abstract}

\maketitle

\section{Introduction}
\label{Introduction}
Precision spectroscopy has a long standing record of providing insights into fundamental physics. In particular, the discrete nature of 
spectral lines has led to the discovery of quantum mechanics and the discovery of the Lamb shift was one of the first confirmations
of quantum electrodynamics. With its ever increasing precision, spectroscopy can continue to provide a powerful probe of new
physics.

Concretely, in this paper we want to use spectroscopy to test Coulomb's law with high precision on atomic length scales~\cite{Popovontheexp,Karshenboim:2010cg, Karshenboim:2010ck}.
This in turn allows us to obtain constraints on new particles such as hidden photons~\cite{Okun,Popovontheexp,Pospelov:2008zw} or minicharged particles~\cite{Gluck:2007ia,Jaeckel:2009dh} which arise naturally in a variety of extensions of the standard model~\cite{Okun,Holdom,Dienes1997104, Abel:2003ue,  Abel:2006qt, Abel:2008ai,Goodsell:2009xc} (see also \cite{Jaeckel:2010ni} for a review)\footnote{Spectroscopy can even be useful for constraining Unparticles (see e.g.~\cite{Thalapillil:2009ch}).}. 

At this point let us note that tests of Coulomb's law are an especially clean and model independent probe of such new particles as the sensitivity does not depend on the stability of the particles or the presence/absence of certain decay channels. 
To illustrate this point, let us take fixed target experiments (see, e.g.~\cite{Bjorken:2009mm}) as an example.
These experiments typically rely on a displaced vertex and therefore on a relatively long decay length of a hidden photon produced in electron-nucleus collision. Now, let us assume that there is also matter charged under the hidden U(1) (i.e., it couples directly to the $B^{\mu}$ field in Eq.~\eqref{LagKM}). If this additional ``hidden matter'' is lighter than $m_{\gamma^{\prime}}/2$, the hidden photon decays quickly into a pair of these particles, dramatically shortening the decay length and possibly invalidating the bound.
Tests of Coulomb's law therefore provide interesting complementary information.

In the following our prime example will be hidden photons which (as we will briefly recall below) cause a deviation of Coulomb's
law of the form
\begin{equation}
V(r)=-\frac{Z \alpha}{r} (1 + e^{-m_{\gamma^{\prime}} r} \chi^2) 
\label{Vintro}
\end{equation}
where $m_{\gamma^{\prime}}$ is the mass of the hidden photon and $\chi$ is a so-called kinetic mixing~\cite{Holdom}.
Independent of this particle interpretation, our bounds can also be taken as a parametrisation of a deviation from
Coulomb's law by a Yukawa type potential with a characteristic length scale $\sim 1/m_{\gamma^{\prime}}$. The length scale of ordinary atoms are typically of the order of the Bohr radius
and correspond to masses of the order of keV. Exotic atoms, such as muonic hydrogen where the electron is replaced by a muon
are even smaller, being most sensitive to masses of order MeV or above. To complement existing spectroscopic tests~\cite{Popovontheexp,Karshenboim:2010cg, Karshenboim:2010ck} we will pay particular attention to the latter regime.

The paper is set up as follows. 
In the next section we will briefly recall how new particles can modify Coulomb's law focusing on our main example, the hidden photon. Minicharged particles are discussed in App.~\ref{Appendix MCP}.
In section \ref{Constraints Using Atomic Spectra} we give a detailed explanation of the method we use to constrain hidden photons, including a discussion of how to obtain the proper behaviour for small (and large) hidden photon masses (section \ref{Renormalisation of alpha}). 
In section \ref{Constructing New Bounds} we apply this method to construct bounds from a variety of atomic spectra. Sections \ref{Bounds from Atomic Hydrogen} and \ref{Ions} deal with ordinary hydrogenic atoms, whereas sections \ref{Exotic Atoms}, \ref{Leptonic Atoms}, and \ref{Muonic Atoms} deal with exotic atoms. 
Particularly interesting speculative bounds are calculated for true muonium and muonic hydrogen.
In sections \ref{Muonium} and \ref{Muonic Atoms} we show that future measurements of true muonium and muonic atoms 
promise significant discovery potential for new physics. Finally, in section \ref{Muonic Hydrogen Anomaly} we discuss whether the hidden photon can be used to explain an anomaly recently observed in muonic hydrogen.
We conclude in section \ref{Conclusion}.

\section{Hidden photons and deviations from Coulomb's law}
\begin{figure}
\begin{center}
\includegraphics[width=145mm]{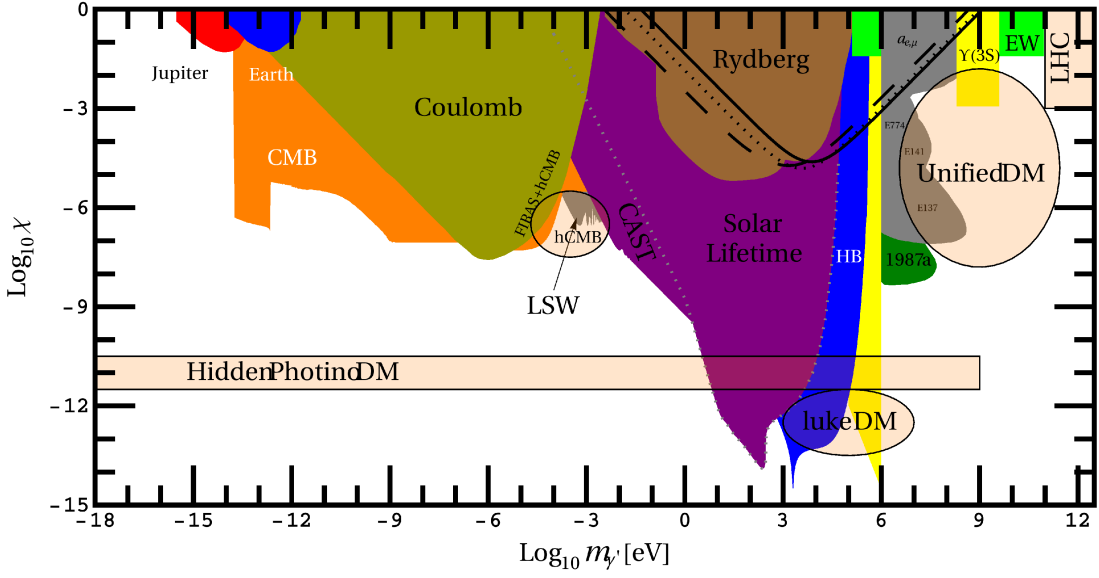}
\caption{Summary of the current bounds on hidden photons (compilation from \cite{Redondoprivate,Jaeckel:2010ni} updated with \cite{Karshenboim:2010cg, Karshenboim:2010ck}). 
We note that in addition to the bound labelled ``Coulomb'', the bounds labelled ``Earth'', ``Jupiter'' and ``Rydberg'' also originate from tests of Coulomb's law.
The best bounds that we derive from atomic spectra are represented by the black lines. The dashed black line is from the  Lamb shift in atomic hydrogen. The dotted black line is a combined bound from the $1s_{1/2} - 2s_{1/2}$ and $2s_{1/2} - 8s_{1/2}$ transitions in atomic hydrogen.
The solid black line is the bound obtained from the Lamb shift in hydrogen-like helium ions.
The light orange areas correspond to regions suggested by astrophysical and cosmological puzzles~\cite{Jaeckel:2008fi, Pospelov:2007mp, Redondo:2008ec, ArkaniHamed:2008qn, Ibarra:2008kn}. The brown region is derived from measurements of the Rydberg constant, and represents bounds already obtained from atomic spectra \cite{Popovontheexp,Karshenboim:2010cg, Karshenboim:2010ck}.
\label{hpcurrent} }
\end{center}
\end{figure}

Hidden photons can interact with the photon via a so-called kinetic mixing~ \cite{Okun,Holdom},
\begin{equation}
L =
-\frac{1}{4}F_{\mu \nu} F^{\mu \nu}
 - \frac{1}{4}B_{\mu \nu} B^{\mu \nu}
+ \frac{\chi}{2} B_{\mu \nu} F^{\mu \nu}
+ \frac{1}{2} m_{\gamma^{\prime}}^2 B_{\mu} B^{\mu}
\label{LagKM}
\end{equation}
where $F_{\mu\nu}$ and $B_{\mu\nu}$ are the photon ($A^\nu$) and hidden photon ($B^\nu$) field strengths. Both $F_{\mu\nu}$ and $B_{\mu\nu}$ are individually gauge invariant, so the third mixing term is allowed. Since the mixing term has a mass dimension of 4, the kinetic mixing parameter $\chi$ has a mass dimension of zero and therefore is a renormalisable coupling. This means that it is not suppressed by any higher mass scales and should be observable (even if small) at lower energies. 
We also note the fourth term which indicates that the hidden photon has a non zero mass.
From the point of view of low energy effective theory, $\chi$ and $m_{\gamma^{\prime}}$ are simply free parameters which are not constrained by any particular physical mechanism. 
However, extensions of the standard model based on, for example, string theory predict values in the 
range~\cite{Dienes1997104,Abel:2008ai,Goodsell:2009xc},
\begin{equation}
10^{-12} \lesssim \chi \lesssim 10^{-3}.
\end{equation}
A summary of the current bounds on the parameter $\chi$ can be found in Fig.~\ref{hpcurrent}.

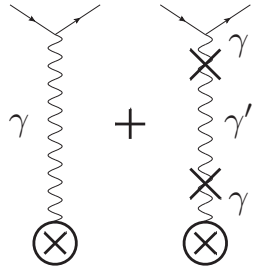
\begin{figure}
\begin{center}
\scalebox{0.35}[0.35]{
 \begin{picture}(258,282) (79,-127)
    \SetWidth{1.0}
    \SetColor{Black}
    \Photon(128,-70)(128,122){7.5}{10}
    \Photon(288,-70)(288,-6){7.5}{3}
    \Photon(288,-6)(288,122){7.5}{6}
    \Text(90,24)[c]{\scalebox{3.5}[3.5]{$\gamma$}}
    \SetWidth{3.0}
    \Line(303.998,-53.998)(272.002,-22.002)\Line(272.002,-53.998)(303.998,-22.002)
    \Line(272.002,74.002)(303.998,105.998)\Line(272.002,105.998)(303.998,74.002)
    \SetWidth{1.0}
    \Photon(128,-70)(128,-102){7.5}{2}
    \Photon(288,-70)(288,-102){7.5}{2}
    \SetWidth{3.0}
    \COval(128,-102)(22.627,22.627)(135.0){Black}{White}\Line(116.686,-90.686)(139.314,-113.314)\Line(139.314,-90.686)(116.686,-113.314)
    \COval(288,-102)(22.627,22.627)(-135.0){Black}{White}\Line(299.314,-90.686)(276.686,-113.314)\Line(299.314,-113.314)(276.686,-90.686)
    \SetWidth{1.0}
    \Line[arrow,arrowpos=0.5,arrowlength=5,arrowwidth=2,arrowinset=0.2](80,154)(128,122)
    \Line[arrow,arrowpos=0.5,arrowlength=5,arrowwidth=2,arrowinset=0.2](128,122)(176,154)
    \Line[arrow,arrowpos=0.5,arrowlength=5,arrowwidth=2,arrowinset=0.2](240,154)(288,122)
    \Line[arrow,arrowpos=0.5,arrowlength=5,arrowwidth=2,arrowinset=0.2](288,122)(336,154)
    \SetWidth{3.0}
    \Line(192.002,26)(223.998,26)\Line(208,41.998)(208,10.002)
     \Text(325,30)[c]{\scalebox{3.5}[3.5]{$\gamma^{\prime}$}}
    \Text(325,110)[c]{\scalebox{3.5}[3.5]{$\gamma$}}
    \Text(325,-60)[c]{\scalebox{3.5}[3.5]{$\gamma$}}
  \end{picture}}
\caption{Feynman diagrams for the potential between two charged particles, Eq.~\eqref{Vcoulomb}. 
In particular the second diagram gives the hidden photon contribution to the interaction between two charged particles, leading
to a modification of Coulomb's law. \label{modified propagators}}
\end{center}
\end{figure}

The kinetic mixing term in Eq.~\eqref{LagKM} causes a tree level insertion to the photon propagator as shown in Fig.~\ref{modified propagators}. An important physical implication of this is the addition of a new Yukawa-type term to the Coulomb potential,
\begin{equation}
V(r)=-\frac{Z \alpha}{r} (1 + e^{-m_{\gamma^{\prime}} r} \chi^2)\equiv V_{\rm Coulomb}(r)+\delta V(r),
\label{Vcoulomb}
\end{equation}
where $Z$ is the charge of the massive central particle which causes the potential and $\alpha$ is the fine structure constant.

Note the following limits,
\begin{itemize}

\item{$m_{\gamma^{\prime}} \rightarrow$ 0}: \\Here the exponential term tends to unity and we recover the original Coulomb potential up to
a factor $(1+\chi^2)$. This can be absorbed in the definition of $\alpha$, making it unobservable.
Physically this is sensible, since in the zero mass limit our hidden photon becomes indistinguishable from the standard model photon, at least as far as the electromagnetic force is concerned. Therefore no physical effects, such as a deviation from Coulomb's law, should be observable.

\item{$m_{\gamma^{\prime}} \rightarrow$ $\infty$}: \\ Here the second exponential term dies off and leaves us with the original Coulomb potential. Again this makes sense, since in the large mass limit the hidden photon becomes impossible to excite as a virtual particle and therefore should not contribute to the electromagnetic force.

\item{Intermediate $m_{\gamma^{\prime}}$}: \\We expect non zero deviations from Coulomb's law in this region. In particular, we will find that the most significant deviation occurs for $m_{\gamma^{\prime}} \sim \frac{1}{l_{0}}$ where $l_{0}$ is the length scale of the relevant physical process.

\end{itemize}

Therefore any constraints on hidden photons based on deviations from Coulomb's law will be strongest around 
$m_{\gamma^{\prime}} \sim \frac{1}{l_{0}}$ and drop off on either side in the low and high mass limits.
Here, $l_{0}$ will be given by the length scale of the atom in question.
There exist a range of different atomic systems, including ordinary hydrogen, muonic atoms, and more exotic atoms. Therefore we can test a wide range of masses, from keV to larger than MeV.

\section{Obtaining constraints using atomic spectra}
\label{Constraints Using Atomic Spectra}
To obtain our constraints we adapt the method presented in Ref.~\cite{Gluck:2007ia}, where the measurement of the Lamb shift is used to derive a bound on minicharged particles. 

At first order in perturbation theory the energy shift of a state $|\psi_{n}\rangle$ is given by
\begin{equation}
\label{1st order}
\delta E_{n}^{(1)} = \langle \psi_{n} \mid H' \mid \psi_{n} \rangle= \langle \psi_{n} \mid \delta V \mid \psi_{n} \rangle
\end{equation}
where the $|\psi_{n} \rangle$ are taken to be the 0th order wave functions.
For this to be a good approximation the energy shift should be small. This is consistent with what we expect, since so far no large deviations from the standard QED predictions have been observed.  
If the standard prediction and the experimentally measured values agree, then we can impose that $\delta E^{(1)}_{n}$ must
be smaller than the uncertainty in the transition. This will result in a bound on $\delta V$.

Let us briefly comment on some points relating to these uncertainties that will be relevant for our discussion.
\begin{itemize}

\item{``Same $n$'' and ``different $n$'' transitions}
\\ We can write the theoretical energy of an arbitrary state as;
\begin{equation}
E_{n,l,j} = E^{D,R}_{n,j} + L_{n,l,j}
\end{equation}
where the first term is the sum of energies from the Dirac equation plus recoil corrections (effectively the 0th order energy). The second term is the Lamb shift (defined as any contribution which separates states of the same $n$,$j$).

The first term is proportional to the Rydberg constant $R_{\infty}$ and therefore will have an uncertainty of approximately $10^{-10}$ eV~\cite{mohr-2007}. This means that any transitions between states of different $n$ in atomic hydrogen will have a theoretical error contribution of around $10^{-10}$ eV from $R_{\infty}$.

The situation is worse in exotic atoms, as the effective Rydberg constant is modified by a factor proportional to the reduced masses $\mu_{H}$, $\mu_{exotic}$ of atomic hydrogen and the exotic atom respectively,
\begin{equation}
R_{eff} = \frac{\mu_{exotic}}{\mu_{H}} R_{\infty}.
\label{Reff}
\end{equation}

\par So that in, for example, muonic hydrogen, there would be a fractional uncertainty of around $10^{-7}$ due to the mass of the muon, which would cause an overall uncertainty of at least $10^{-4}$ eV.
\par Note that transitions between same $n$ states do not have a 0th order energy and are limited only by uncertainties in the Lamb shifts of the states.

\item{Definition of ``uncertainty''} \\ 

To be very conservative when forming bounds, we estimate the total uncertainty of a transition by adding together the absolute values of the theoretical and experimental errors, i.e. for a given measurement $M$ we use
\begin{equation}
\Delta M = |\Delta M (th)| + |\Delta M (exp)|
\label{deltanoH}
\end{equation}

The theoretical contribution to $\Delta M$ will come mainly from uncertainties in the finite nuclear size, with an additional contribution from the Rydberg constant in the case of different $n$ transitions. For most of the atoms we consider, data for the finite nuclear size is coherent. For example the hydrogenlike helium ion has three different electron scattering determinations, and also a muonic helium ion determination, of the alpha particle charge radius. These values all agree within 1 $\sigma$ \cite{PhysRevA.43.3325}, meaning that to a good approximation we can calculate the theoretical value for a transition by assuming one particular value of the nuclear radius. We can estimate the uncertainty from finite nuclear size effects simply by considering the uncertainty in this one value of the nuclear radius. This is indeed what is done in theoretical calculations to which we refer, and the corresponding error is included in $|{\Delta M}_{th}|$.

However when we consider atoms with a proton nucleus things become more complicated. The recent muonic hydrogen determination of $r_{p} = 0.84184(67)$ fm \cite{citeulike:7426442} gives us the most precise measurement from atomic spectra. This disagrees with our best previous atomic spectra determination of $r_{p} = 0.8768(69)$ fm from \cite{mohr-2007} by around 5 $\sigma$. The muonic hydrogen extraction also deviates by around 2.5 $\sigma$ from the best electron scattering determination of $r_{p} = 0.897(18)$ fm \cite{PhysRevC.72.057601}. 
To be conservative we therefore modify our error analysis to take into account the large variation in $r_{p}$. We do this by 
adding an additional term $| \Delta M (r_{p})|$, accounting for the discrepancies in the $r_{p}$ measurements,
\begin{equation}
\Delta M^{*} = |\Delta M (th)| + |\Delta M (exp)| + | \Delta M (r_{p})|.
\label{deltaH}
\end{equation}
To leading order the proton radius contribution to a given state is 
\begin{equation}
E_{NS} ( r_{p} ) = \frac{2 m_{o}^3 \alpha^4 r_{p}^2}{3 n^{3}}\delta_{l0},
\label{rpdep}
\end{equation}
where $m_{o}$ is the mass of the orbiting particle. In atomic hydrogen $m_{o} = m_{e}$, and in muonic hydrogen $m_{o} = m_{\mu}$. 
We make a rough estimate for $| \Delta M (r_{p})|$ by considering two widely separated values of the proton radius. Denoting $r_{p, \mu} =0.84184(67)$ fm as the muonic hydrogen determination, and $r_{p,e} = 0.897(18)$ fm as the electron scattering determination, we set
\begin{equation}
| \Delta M (r_{p})| = | E_{NS} (r_{p,e}) - E_{NS} (r_{p,\mu}) |.
\label{rperror}
\end{equation}

To form bounds we use $\Delta M^{*}$ for atoms with a proton nucleus, which in our paper are atomic hydrogen and muonic hydrogen. We use $\Delta M$ for all other atoms. 

Note that errors quoted in the text will always be at the 1 $\sigma$ ($\sim 68\%$ confidence level). However, unless otherwise stated we will consider $2\sigma$ errors ($\sim 95\%$ confidence level) for the bounds produced in the figures. This essentially means that all terms on the right hand side of \eqref{deltaH}, \eqref{deltanoH} are multiplied by a factor of 2.

\item{Radius of the proton and other nuclei\\
As already discussed the (charge) radii of the nuclei are a major source of uncertainty. In addition, we have to take care that our determination of
the radii is from an independent source. For example we can \emph{not} take a measurement of the Lamb shift in ordinary hydrogen to measure both
the radius of the proton and put a bound on deviations from Coulomb's law. Two independent measurements are needed. Moreover
to avoid even partial degeneracies (which tend to weaken the bound in particular at short length scales), it is best if the determination of the radius is obtained at relatively high momentum transfer, corresponding to a short length scale. For this reason we will mainly use the values obtained from electron scattering data.}

\end{itemize}

\subsection{Searching for deviations from Coulomb's law}
\label{Renormalised Method for chi}
\subsubsection{Naive bounds}
\label{Naive Bounds for chi}
Let us briefly demonstrate how this method works by using the Lamb shift, i.e. the $2s_{1/2}-2p_{1/2}$ transition in atomic hydrogen to constrain the hidden photon. We use $\delta V$ given by Eq.~\eqref{Vcoulomb}. 

For this case Eq.~\eqref{1st order} evaluates,
\begin{equation}
\delta E = \int_0^{\infty} dr\, r^2\delta V(r) 
  \left[R_{20}^2(r) - R_{21}^2(r)\right]=-\chi^2\frac{\alpha \, a \, m^{2}_{\gamma^{\prime}}}{2 (1+am_{\gamma^{\prime}})^4}.
\label{lambshift}
\end{equation}
where we have used the normalised radial hydrogen wave functions $R_{n\ell}$ given by
\begin{equation}
R_{20}(r) = \frac{1}{\sqrt{2}}\,\, \frac{1}{a^{3/2}}
   \left( 1-\frac{\rho}{2}\right) \exp(-\rho/2)\,, \quad
R_{21}(r) = \frac{1}{2\sqrt{6}}\,\,\frac{1}{a^{3/2}}\rho\, \exp(-\rho/2)
\end{equation}
and where $\rho= r/a$ and $a^{-1}=\alpha m_e$.

\begin{figure}
\begin{center}
\includegraphics[width=100mm]{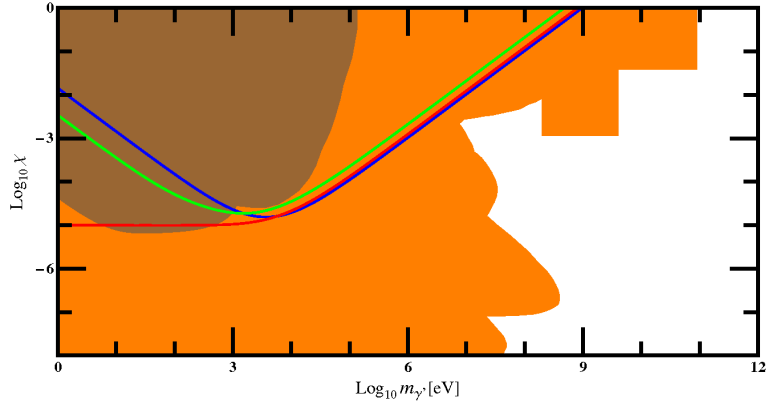}
\caption{The blue curve denotes the bound on hidden photons obtained from the $2s_{1/2} - 2p_{1/2}$ transition in atomic hydrogen. We use a conservative error of $\Delta M^{*} = 2 \times 10^{-10}$ eV at the 2 $\sigma$ level where $\Delta M^{*}$ is defined in \eqref{deltaH}. Note that for same $n$ transitions $\chi$ dies off correctly for both small and large $m_{\gamma^{\prime}}$. The red curve shows the naive bound from the $1s_{1/2} - 2s_{1/2}$ transition in atomic hydrogen. It has an incorrect behaviour at small $m_{\gamma^{\prime}}$. The green curve gives the correct bound obtained by combining the $1s_{1/2} - 2s_{1/2}$ with the $2s_{1/2} - 8s_{1/2}$ transition, according to the procedure described in Sect.~\ref{Renormalisation of alpha}. 
The green and blue bounds will turn out to be the best ones that we can derive from atomic spectra, and are combined together in Fig. \ref{hpcurrent}.
For comparison we depict earlier bounds on hidden photons as the colour filled regions. Those corresponding to pure tests of Coulomb's law (also obtained from atomic spectra) are highlighted in brown \cite{Popovontheexp,Karshenboim:2010cg, Karshenboim:2010ck}. The remaining white region corresponds to unexplored parameter space. 
\label{combinedhydrogen}}
\end{center}
\end{figure}

We use an experimental uncertainty of 3 kHz \cite{PhysRevLett.82.4960} and theoretical uncertainty of 6 kHz \cite{Simon:1980hu}, as well as a contribution of 17 kHz from $| \Delta M (r_{p})|$ \eqref{rperror}. We use \eqref{deltaH} to get $\Delta M^{*} = 10^{-10}$ eV at the  1 $\sigma$ level. Note that the (blue) curve for the $2s_{1/2}-2p_{1/2}$ transition (Fig.~\ref{combinedhydrogen}) is at the 2 $\sigma$ level.

This has the correct shape; the bound dies off the in the limits $m_{\gamma^{\prime}} \rightarrow$ 0 and $m_{\gamma^{\prime}} \rightarrow \infty$, and is strongest at $m_{\gamma^{\prime}} \sim \frac{1}{a}$, where the Bohr radius $a$ is the typical length scale involved. 

We can do the same for the $1s_{1/2}-2s_{1/2}$ transition in atomic hydrogen.
The experimental value has a relative uncertainty $2.8 \times 10^{-14}$ and represents the most precise measurement of atomic 
hydrogen~\cite{PhysRevLett.92.230802}.
However, the bounds are limited by a much larger theoretical uncertainty. As already discussed above there are uncertainties of around $10^{-10}$ eV from the Rydberg constant as well as a similar contribution from the Lamb shifts of the states \cite{Biraben}. We also need to add the $| \Delta M (r_{p}) |$ contribution of $\sim 7 \times 10^{-10}$ eV to get $\Delta M^{*} = 1 \times 10^{-9}$ eV at the 1 $\sigma$ level.

For the first order energy shift we have
\begin{eqnarray}
\delta E& = &\int_0^{\infty} dr\, r^2\delta V(r) 
  \left[R_{20}^2(r) - R_{10}^2(r)\right]
  \\\nonumber
  &=&\chi^2\alpha\left[\frac{12+am_{\gamma^{\prime}}(60+am_{\gamma^{\prime}}(87+14am_{\gamma^{\prime}}(4+am_{\gamma^{\prime}})))}{4a(1+am_{\gamma^{\prime}})^4(2+am_{\gamma^{\prime}})^2}\right],
\end{eqnarray}
where
\begin{equation}
R_{10}(r) = \frac{2}{a^{3/2}}\exp(-\rho).
\end{equation}

The result is shown in Fig.~\ref{combinedhydrogen} as the red curve. This bound does not have the correct drop off for small masses. We can understand why by looking at $\delta V(r)$ in Eq.~\eqref{Vcoulomb}, which dies off at large masses but grows at small masses. Therefore we get a bound which saturates at small masses, which is not physically correct. This is simply an artefact of the splitting of the potential that we have chosen to set up our perturbation theory. At small masses our perturbation reduces to a term that has the form of a Coulomb potential, but with an extra factor $(1+\chi^2)$. 
This effectively increases the strength of the electromagnetic coupling and therefore the energy difference between the two states. In other words we have forgotten to properly (re-)normalise the coupling $\alpha$, i.e. we have not absorbed the factor $(1+\chi^2)$ into our definition of $\alpha$.
In the following we will show explicitly how this can be remedied.

One might wonder why this problem does not affect the Lamb shift bound. The reason is simply that for a perfect Coulomb's law the energies
of $2s_{1/2}$ and the $2p_{1/2}$ are degenerate. Therefore, adding a term of the Coulomb's law form does not produce an energy shift between the two states at lowest order. This is true for all same $n$ transitions.

\subsubsection{Bounds including a proper renormalisation of $\alpha$}
\label{Renormalisation of alpha}

As we have seen above, the problem is that for both $m_{\gamma^{\prime}}\to\infty$ and $m_{\gamma^{\prime}}\to 0$
the potential Eq.~\eqref{Vcoulomb} has exactly the $1/r$ behaviour of a Coulomb potential. However the coupling constant
differs by a factor $(1+\chi^2)$.
$\alpha$ becomes a function of $\chi$ and $m_{\gamma^{\prime}}$, i.e the hidden photon alters the coupling constant itself. In that sense $\alpha$ becomes an unknown quantity that needs to be fixed by experiment. Since we now have two unknowns, $\chi$ and $\alpha$, we need two measurements to solve for them (of course the same strategy also works if we allow other/additional parameters to vary). 
We will briefly sketch how we do this.

Let us assume we have two observables $M_{1}$ and $M_{2}$. 
Theoretically these are functions of $\alpha$, $\chi^2$ and $m_{\gamma^{\prime}}$. To keep the notation transparent we will suppress the dependence on $m_{\gamma^{\prime}}$ in the following. Therefore, we have $M_{1}(\alpha,\chi^2)$ and $M_{2}(\alpha,\chi^2)$.
 
Now we have two measurements and results are often quoted in the form
\begin{eqnarray}
M_{1}|_{\rm exp}-M_{1}|_{\rm th}=\delta M_{1}\pm \Delta M_{1},
\\
M_{2}|_{\rm exp}-M_{2}|_{\rm th}=\delta M_{2}\pm \Delta M_{2},
\end{eqnarray}
without considering a hidden photon. Therefore, in our setup this means 
\begin{eqnarray}
M_{1}|_{\rm exp}-M_{1}(\alpha_{0},0)=\delta M_{1}\pm \Delta M_{1},
\\
M_{2}|_{\rm exp}-M_{2}(\alpha_{0},0)=\delta M_{2}\pm \Delta M_{2},
\end{eqnarray}
with some value $\alpha_{0}$ such that both $\delta M_{1}$ and $\delta M_{2}$ are small. If $\alpha_{0}$ can be chosen that $\delta M_{1}$ and $\delta M_{2}$ are compatible with $0$ within the errors, then the measurements are consistent with the standard model and no hidden photon is required.

Now, what happens if we include a hidden photon? As we are interested in small corrections, we can expand about 
$(\alpha,\chi)=(\alpha_{0},0)$. Hence we obtain
\begin{eqnarray}
\frac{\partial M_{1}}{\partial{\alpha}}|_{\alpha=\alpha_{0},\chi^2=0} \,\delta \alpha+\frac{\partial M_{1}}{\partial{\chi^2}}|_{\alpha=\alpha_{0},\chi^2=0} \,\chi^2=\delta M_{1}\pm \Delta M_{1},
\\
\frac{\partial M_{2}}{\partial{\alpha}}|_{\alpha=\alpha_{0},\chi^2=0}\delta \alpha+\frac{\partial M_{2}}{\partial{\chi^2}}|_{\alpha=\alpha_{0},\chi^2=0}\chi^2=\delta M_{2}\pm \Delta M_{2}.
\end{eqnarray}
In matrix notation this linear system of equations reads,
\begin{equation}
\left(
\begin{array}{cc}
\frac{\partial M_{1}}{\partial{\alpha}}|_{\alpha=\alpha_{0},\chi^2=0}  &    \frac{\partial M_{1}}{\partial{\chi^2}}|_{\alpha=\alpha_{0},\chi^2=0}  \\
  \frac{\partial M_{2}}{\partial{\alpha}}|_{\alpha=\alpha_{0},\chi^2=0}&      \frac{\partial M_{2}}{\partial{\chi^2}}|_{\alpha=\alpha_{0},\chi^2=0}
\end{array}
\right)
\left(
\begin{array}{c}
  \delta \alpha \\
    \chi^2
\end{array}
\right)
=
\left(
\begin{array}{c}
  \delta M_{1}\pm \Delta M_{1} \\
  \delta M_{2}\pm \Delta M_{2}
\end{array}
\right)
\end{equation}
which can be easily solved
\begin{equation}
\left(
\begin{array}{c}
  \delta \alpha \\
    \chi^2
\end{array}
\right)
=
\left(
\begin{array}{cc}
\frac{\partial M_{1}}{\partial{\alpha}}|_{\alpha=\alpha_{0},\chi^2=0}  &    \frac{\partial M_{1}}{\partial{\chi^2}}|_{\alpha=\alpha_{0},\chi^2=0}  \\
  \frac{\partial M_{2}}{\partial{\alpha}}|_{\alpha=\alpha_{0},\chi^2=0}&      \frac{\partial M_{2}}{\partial{\chi^2}}|_{\alpha=\alpha_{0},\chi^2=0}
\end{array}
\right)^{-1}
\left(
\begin{array}{c}
  \delta M_{1}\pm \Delta M_{1} \\
   \delta M_{2}\pm \Delta M_{2}
\end{array}
\right).
\label{general}
\end{equation}
From this we can directly read off the allowed values of $\chi^2$. 

For the cases of interest to us it is typically sufficient to determine the derivatives $\partial M_{i}/\partial \alpha$, $\partial M_{i}/\partial \chi^2$
to lowest order in $\alpha$ and $\chi^2$. For example, in the previous subsection we have used first order perturbation theory, and determined
the shift in energy to be $\delta E=\chi^2 f(m_{\gamma^{\prime}})$, accordingly $\partial E /\partial \chi^2=f(m_{\gamma^{\prime}})$.

Let us confirm that this procedure corrects the behaviour of bounds for small $m^{2}_{\gamma^{\prime}}$. We consider a simple example
where, in absence of hidden photons, both observables behave as simple power laws, $\sim \alpha^{n_{1}}$ and $\sim \alpha^{n_{2}}$, respectively,
\begin{equation}
M_{i} = M_{i}|_{\chi^2=0} + \delta_{\chi^2} M_{i} = c_{i} \alpha^{n_{i}} + c_{i} \alpha^{n_{i}} \chi^2 f_{i}(m_{\gamma^{\prime}}) + O(\chi^4).
\label{M1}
\end{equation}
The term in $\delta_{\chi^2} M_{i}$ is the hidden photon contribution calculated in first order perturbation theory~\eqref{1st order} using the 
potential~\eqref{Vcoulomb}. Note that for convenience we have written the correction with $c_{i}$ and $\chi^2$ factored out, with the remaining factor represented by a function $f_{i}(m_{\gamma^{\prime}})$.

From Eq.~\eqref{Vcoulomb} we can see that $\delta V(r)$ term dies off at large $m_{\gamma^{\prime}}$. Hence the function $f_{i}(m_{\gamma^{\prime}}) \rightarrow 0$ as $m_{\gamma^{\prime}} \rightarrow \infty$. However, the perturbation actually grows towards smaller masses, and $f_{i}(m_{\gamma^{\prime}})$ tends to a constant limit.
Before addition of the hidden photon we can write the Coulomb potential as
\begin{equation}
V_{0}(r) = - \frac{Z \alpha}{r}
\label{originalCoulomb}
\end{equation}
and after the hidden photon is added we can write
\begin{equation}
V(r) = - \frac{Z \alpha}{r} (1 + \chi^2)
\label{VIR}
\end{equation}
where we have taken the limit $m_{\gamma^{\prime}} \rightarrow 0$.
Hence the fine structure constant has essentially been redefined
\begin{equation}
\alpha \rightarrow \alpha (1 + \chi^2).
\label{alpharedefined}
\end{equation}

Therefore, for $m_{\gamma^{\prime}}\to 0$, we have an alternative way to obtain the perturbation: we can simply insert the redefined $\alpha$
into the unperturbed expression.
This yields,
\begin{equation}
M_{i} = c_{i} \alpha^{n_{i}} + n_{i} c_{i} \alpha^{n_{i}} \chi^2 + O(\chi^4).
\label{M10-M1}
\end{equation} 
Comparing \eqref{M1} with \eqref{M10-M1} we can see that $f_{i}(m_{\gamma^{\prime}}) \rightarrow n_{i}$ as $m_{\gamma^{\prime}} \rightarrow 0$.

Inserting \eqref{M1} into our general expression \eqref{general} we find,
\begin{equation} \chi^2 = \frac{\frac{n_{1} \delta M_{2}}{M_{2}(\alpha_{0},0)} - \frac{n_{2} \delta M_{1}}{M_{1}(\alpha_{0},0)}} {(n_{1} f_{2}(m_{{\gamma^{\prime}}}) - n_{2} f_{1}(m_{{\gamma^{\prime}}}))}.
\end{equation}

If both $\delta M_{i}$ are consistent with $0$ we can obtain a bound on $\chi^2$. To be conservative we simply add the moduli of the two individual errors,
\begin{equation}\chi^2 \leq \frac{\frac{n_{1} | \Delta M_{2} | } {M_{2}} + \frac{n_{2} | \Delta M_{1} | }{M_{1}}} {|(n_{1} f_{2}(m_{{\gamma^{\prime}}}) - n_{2} f_{1}(m_{{\gamma^{\prime}}}))|}.
\label{chiren}
\end{equation}
where $| \Delta M |$ is defined in \eqref{deltanoH}. For transitions in atomic hydrogen and muonic hydrogen we use $| \Delta M^{*} |$ defined in \eqref{deltaH}.

There are two interesting limits\footnote{We note, that in the regions where the bounds become very weak and $\chi={\mathcal O}(0.1-1)$, perturbation theory breaks down. But these regions are typically excluded anyway.}
\begin{itemize}
\item{$m_{\gamma^{\prime}} \rightarrow \infty$}: Here $f_{1}(m_{\gamma^{\prime}}), f_{2}(m_{\gamma^{\prime}}) \rightarrow 0$ so the upper bound 
on $\chi^2$ increases and the bound dies off.
Indeed one finds that the functions $f(m_{\gamma^{\prime}})$ decay as $m_{\gamma^{\prime}}^{-(2 + 2l)}$ at high masses where $l$ is the lowest angular momentum value involved in the transition. A $2s - 2p$ transition would have $l = 0$ and a $3p - 3d$ transition would have $l = 1$ etc. Our bounds effectively die off as 
\begin{equation}
\chi \sim m_{\gamma^{\prime}}^{(1 + l)}.
\label{powerlawdecay}
\end{equation} 
Therefore, if we use transitions with higher values of $l$ the bounds die off more quickly.

\item{$m_{\gamma^{\prime}} \rightarrow 0$}: Here $f_{1}(m_{\gamma^{\prime}}) \rightarrow n_{1}, f_{2}(m_{\gamma^{\prime}}) \rightarrow n_{2}$ so the denominator tends to zero. Again the upper limit on $\chi^2$ increases and the bound dies off.
\end{itemize}
Overall, we obtain the expected behaviour in the small and large mass limits.

In Fig.~\ref{combinedhydrogen} we show the correctly renormalised $1s_{1/2} - 2s_{1/2}$ (green) versus the naive bound (red).
Here, and unless otherwise stated we will take $M_{1}$ to be the $2s_{1/2} - 8s_{1/2}$ transition in atomic hydrogen, simply because it is experimentally measured to a high precision of $3 \times 10^{-11}$~eV \cite{PhysRevLett.78.440}. 
This is similar to the theoretical error of $4.5 \times 10^{-11}$~eV from $R_{\infty}$ and $5 \times 10^{-11}$~eV from the Lamb shift of the $2s_{1/2}$ state. Added together with the $\Delta M (r_{p})$ contribution we get an overall uncertainty of $2 \times 10^{-10}$~eV at the 1 $\sigma$ level.

The measurements themselves do not need to be from atomic spectra. The same technique works for any process which is affected by the hidden photon. 

Also note that this renormalisation procedure works trivially for same $n$ transitions. If we consider just one measurement $M$, and note that the 0th order energy $M_{0}$ vanishes,
\begin{equation}
M = c \alpha^{2} \chi^2 f_{2}(m_{\gamma^{\prime}}) + O(\chi^4).
\end{equation}
$M \propto \chi^{2}$, so that $\delta \alpha$ corrections obtained from a second measurement would produce negligible terms of O($\chi^4$). Therefore information from the second measurement is suppressed. This means that we can form properly renormalised bounds for same $n$ transitions using only one measurement $M$ and with only one error $\Delta M$.

\section{New bounds}
\label{Constructing New Bounds}

\subsection{Ordinary atomic hydrogen}
\label{Bounds from Atomic Hydrogen}
The bounds from $2s_{1/2} - 2p_{1/2}$ (blue) and the properly renormalised $1s_{1/2} - 2s_{1/2}$ bound (green) are shown in Fig.~\ref{combinedhydrogen}. We can see that neither of these penetrate new parameter space for hidden photons. However, they constitute the best pure tests of Coulomb's law in this region of parameter space. 
 
So far, all of our transitions have involved the $1s_{1/2}$ or $2s_{1/2}$ states. These have Lamb shifts with high theoretical uncertainty due to finite nuclear size effects. This is because the 0th order wavefunctions for $s$ states are non zero at the origin and therefore penetrate the nucleus deeply, leading to large uncertainties from finite nuclear size effects. This will also be true for $s$ states with $n \, \textgreater \, 2$. 
States with $l \, \textgreater \, 0$ have 0th order wavefunctions which are zero at the origin and therefore have a small overlap with the nucleus. This means that the Lamb shift uncertainties are small. Therefore transitions between same $n$ states with $l \, \textgreater \, 0$ (for example $3p_{1/2} - 3d_{1/2}$) will have extremely small theoretical uncertainties. Unfortunately, these kinds of excited states are unstable and the experimental measurements have large uncertainties. We also note that the bounds in these transitions decay more quickly with energy \eqref{powerlawdecay}, causing them to die off well before they reach unknown parameter space. Therefore, barring significant technological advances, transitions between states with $l \, \textgreater \, 0$ are not very useful for constraining $\chi$.

\subsection{Ions with $Z \, \textgreater \, 1$}
\label{Ions}
The Lamb shift bound for atomic hydrogen almost penetrates the unknown region, so we can look at hydrogenic ions with $Z \textgreater 1$. The advantage here is that the characteristic energy of transitions scales as $Z$, so higher values of $Z$ will move our bounds towards the right and towards the white region. 
The disadvantage is the higher theoretical uncertainties involved. This is due to an increase in the size of the nucleus, as well as a decrease in the Bohr radius (which scales as $ \frac{1}{Z} $ and gives the characteristic length scale of the electron orbit). 
Unfortunately, this also causes the electron to penetrate the nucleus more deeply, which leads to greater theoretical uncertainties from finite nuclear size effects. Our bounds then move upwards and away from the white region.
We can produce bounds for different values of $Z$ to see which of these effects wins out. 

\begin{figure}
\begin{center}
\includegraphics[width=100mm]{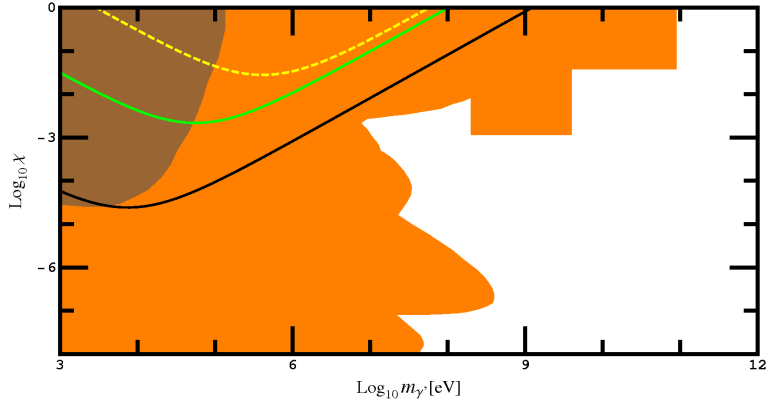}
\caption{$\chi$ bounds for the $2s_{1/2}-2p_{1/2}$ transition in hydrogenic ions with $Z = 2$ (black), $Z = 15$ (green), and $Z = 110$ (yellow dashed, speculative). At the 2 $\sigma$ level the uncertainties are $3 \times 10^{-9}$ eV \cite{PhysRevA.43.3325}, $6 \times 10^{-4}$ eV \cite{0295-5075-5-6-005, Johnson1985405} and 8 eV \cite{Johnson1985405} respectively. We can see that as $Z$ increases our bounds move away from the white region and therefore become less useful. \label{diffZ}}
\end{center}
 
\end{figure}

Ref. \cite{PhysRevA.43.3325} gives experimental and theoretical errors for the helium ion $2s_{1/2} - 2p_{1/2}$ transition of 0.16 MHz and 0.2 MHz respectively, giving $\Delta M = 1.5 \times 10^{-9}$ eV at the 1 $\sigma$ level. (Remember that for atoms without a proton nucleus, we do not have to consider large variations in nuclear size, and therefore use \eqref{deltanoH}.) For $Z = 15$ the experimental~\cite{0295-5075-5-6-005} and theoretical 
errors~\cite{Johnson1985405} combine to give a value of $\Delta M = 3 \times 10^{-4}$ eV at the 1 sigma level. Finally we can go to the largest value of $Z$ for which data is available. Ref. \cite{Johnson1985405} gives a theoretical uncertainty for the $Z = 110$ transition of $\sim$ 4 eV. 
As no experimental data exists we can only derive a speculative bound.
Also, if we were to take into account hypothetical experimental data, then the errors would increase and the bound would weaken. The estimated sensitivity shown in yellow in Fig.~\ref{diffZ} is therefore an optimistic one, but it is still adequate to demonstrate the trend which we are trying to identify.  

We can see that as we go to higher values of $Z$, the increase in uncertainties cause our bounds to move up and away from the untested region, with our bounds actually becoming less and less stringent. We should expect this trend to be similar for other possible transitions, for example $1s_{1/2}-2s_{1/2}$.

Note that due to the increase in the uncertainty of the Lamb shift in atomic hydrogen caused by the inconsistent values of the proton radius, the bound from hydrogen-like helium is actually slightly better. As soon as this inconsistency is clarified the trend identified above will probably hold again. 

\subsection{Exotic atoms}
\label{Exotic Atoms}
Exotic atoms have certain advantages over atomic hydrogen:
\begin{itemize}
\item{Pure QED systems like muonium and positronium may have smaller fractional theoretical uncertainties, as all experimental data is consistent with pointlike leptons \cite{Jungmann:2004sa}. This enables us to assume a pointlike effective nucleus, which eliminates the major source of theoretical uncertainty.}
\item{Most exotic atoms have larger reduced masses than atomic hydrogen, shifting our bounds to higher energies and towards the untested region.}
\end{itemize}

Disadvantages:
\begin{itemize}
\item{Higher mass systems often have greater theoretical uncertainty (leptonic systems excluded). This is because the larger reduced mass leads to a smaller Bohr radius of the system. Bound state orbits then penetrate the nucleus more deeply, which then leads to greater corrections due to finite nuclear size effects.} 
\item{Similarly, hadronic atoms and atoms with $Z \textgreater 1$ have larger nuclei than atomic hydrogen, again causing larger finite nuclear size effects.}
\item{Hadronic orbiting particles also interact with the nucleus via the strong interaction, which causes huge theoretical and experimental uncertainties.They do not produce strong bounds, but for completeness we briefly review them in App.~\ref{Appendix Hadronic}.}

\end{itemize}

\subsection{Leptonic atoms}
\label{Leptonic Atoms}

\subsubsection{Positronium}
\label{Positronium}

We briefly note that positronium is not useful as the reduced mass is actually smaller than that of atomic hydrogen, and the uncertainties are much higher. For example the $1s - 2s$ and $2s - 2p$ transitions are limited by large experimental uncertainties, which are caused by complications such as annihilation \cite{PhysRevLett.71.2887, springerlink:10.1007/BF02064517 ,Ley2002301}. The uncertainties are around two orders larger, which gives a bound on $\chi$ around one order of magnitude weaker. 

\subsubsection{Muonium}
\label{Muonium}

Ordinary muonium ($\mu^{+}$ $e^{-}$) and true muonium ($\mu^{+}$ $\mu^{-}$) are more interesting.

Experimental results have already been produced for the $1s_{1/2} - 2s_{1/2}$ and $2s_{1/2} - 2p_{1/2}$ transitions in ordinary muonium, but the resulting bounds do not improve on atomic hydrogen. We note that the reduced mass of ordinary muonium is almost the same (in fact slightly smaller) than atomic hydrogen, so we could only get better bounds if uncertainties were reduced.

In fact the $1s_{1/2} - 2s_{1/2}$ transition suffers from large theoretical errors associated with the effective Rydberg constant~\eqref{Reff}. 
This produces a fractional error larger than that of atomic hydrogen and consequently a weaker bound.

The $2s_{1/2} - 2p_{1/2}$ transition suffers from no such 0th order uncertainties, and indeed we expect the theoretical uncertainty to be smaller than atomic hydrogen since there are no finite nuclear size effects. However the experimental situation is not yet very good. Only $1s_{1/2}$ states can be produced in large quantities, whereas the production of metastable $2s_{1/2}$ states is much lower \cite{jungmann-1999}. This means that the $2s_{1/2}-2p_{1/2}$ transition has only been experimentally measured to a fractional precision of $1.5 \times 10^{-2}$ at the 1 $\sigma$ level, leading to weak bounds.

\begin{figure}
\begin{center}
\includegraphics[width=100mm]{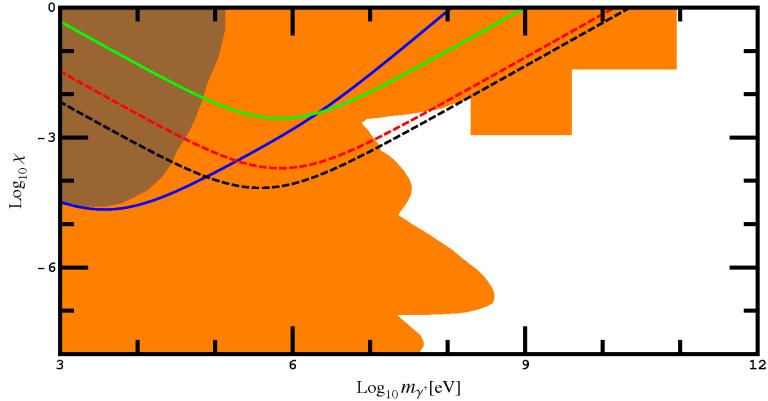}
\caption{The black dashed curve shows a speculative bound on $\chi$ for the $2s_{1/2}-2p_{1/2}$ transition in true muonium. This is formed using only an estimate of the theoretical uncertainty of $\sim 0.1$ GHz (see appendix \ref{Appendix Muonium Estimate}). If the experimental result is measured to a similar precision and agrees with theory, we will be able to form a bound similar to this and penetrate new parameter space.
The green solid curve shows an actual bound formed from the $2s_{1/2}^{F=1} - 2p_{3/2}^{F=2}$ transition in muonic hydrogen. There is a large theoretical error $\Delta M^{*} = 1.2\times 10^{-3} $~eV at the 2 $\sigma$ level. This is caused by the wide range of possible values of $r_{p}$.
The red dashed curve uses the muonic hydrogen transition to form a speculative bound. The error here is taken to be just the 2 $\sigma$ experimental uncertainty of $6 \times 10^{-6}$~eV \cite{citeulike:7426442}. Finally, the blue curve shows a bound obtained by combining the measurement of the Lamb shift in ordinary hydrogen and the $2s_{1/2}^{F=1} - 2p_{3/2}^{F=2}$ transition in muonic hydrogen. Using these two measurements we do not need an additional determination of the proton radius. 
\label{specbounds}}
\end{center}
\end{figure}

A recent article \cite{Brodsky:2009gx} suggests that true muonium can be produced and studied in the near future. This atomic system would be extremely useful, as it has a reduced mass of around 2 orders of magnitude greater than atomic hydrogen, and we expect the theoretical errors to be low due to the absence of finite nuclear size effects.
Since no experimental data has been produced yet, there has been no motivation for precise theoretical calculations. However, we can put together a rough theoretical estimate (see App.~\ref{Appendix Muonium Estimate} for details) to form a speculative bound. This is shown as the black line in Fig.~\ref{specbounds}. This is encouraging as it penetrates new parameter space. 
However, one still needs to obtain a coherent experimental result, and hope that the experimental error is not so large that it causes the bound to significantly deteriorate.
 
\subsection{Muonic atoms}
\label{Muonic Atoms}
The reduced mass of these systems is around 200 times larger than atomic hydrogen, shifting our bounds towards larger masses.

The $2s_{1/2}^{F=1} - 2p_{3/2}^{F=2}$ difference in muonic hydrogen is calculated to be \cite{Eides:2000xc,PhysRevA.53.2092,PhysRevA.60.3593,PhysRevA.71.032508,PhysRevA.71.022506,hypfinesplit,citeulike:7426442}
\begin{equation}
E =-\left[209.9779(49) - 5.2262 \left(\frac{r_{p}}{\rm fm}\right)^2 + 0.0346 \left(\frac{r_{p}}{\rm fm}\right)^{3}\right] \, \, \text{meV}.
\label{muonichth}
\end{equation}

If we substitute in the most precise current value of $r_{p} = 0.8768(69)$ fm which is obtained from atomic spectra \cite{mohr-2007}, we obtain a theoretical value of $E = -205.984(062)$ meV. The theoretical uncertainty alone is quite high. 
Moreover, the theoretical value also deviates from the recently measured experimental value of $-206.295000(3)$ meV \cite{citeulike:7426442} by around 5$\sigma$. This large discrepancy is bad for producing bounds, but it could be taken as a potential signal for new physics. We will consider this in the final Sect.~\ref{Muonic Hydrogen Anomaly}.
However, we can still form a bound if we use the (inflated) error given in \eqref{deltaH}. The $r_{p}$ variation increases the uncertainty to give 
$\Delta M^{*} = 6\times 10^{-4}$~eV at the 1 $\sigma$ level. The solid green curve shows the corresponding bound at the 2 $\sigma$ level. 
In the large mass region this bound is of comparable strength to the one from the Lamb shift in ordinary hydrogen.

However if we just use the experimental uncertainty we can form a speculative bound Fig.~\ref{specbounds}. This bound covers a similar region to the speculative bound obtained from true muonium (black curve) and penetrates the unknown region.

If an independent and sufficiently precise value of $r_{p}$ could be determined -- consistent with the muonic hydrogen extraction --
the speculative bound could be turned into a real one.
This provides us with motivation for seeking more precise, independent determinations of the proton radius.
However, as we have discussed in Sect.~\ref{Constraints Using Atomic Spectra}, in order to avoid degeneracies this measurement should preferably originate from measurements at relatively high momentum transfer.
At the moment the only obvious candidate process is electron scattering, although an increase in the precision by an order of magnitude may be challenging.

\section{Muonic hydrogen anomaly$^{*}$}
\label{Muonic Hydrogen Anomaly}
\renewcommand{\thefootnote}{\fnsymbol{footnote}}
\footnotetext[1]{We are deeply indebted to B.~Batell and M.~Pospelov for noting a sign error in a previous version of this paper which led to different results. In the following discussion of the effects of hidden photons on measurements of the proton radius we will use an argument similar to theirs.}
\renewcommand{\thefootnote}{\arabic{footnote}}

As already mentioned above a recent measurement~\cite{citeulike:7426442} of the Lamb shift in muonic hydrogen, or more precisely the $2s_{1/2}^{F=1} - 2p_{3/2}^{F=2}$ transition, deviates by more than $5\sigma$ from theoretical calculations combined with atomic spectra
measurements of the proton radius.
It is tempting to speculate that this deviation is due to a hidden photon~\cite{falkowskidark}.
In this section we will briefly investigate if such an interpretation is possible.

The first observation is that the addition of the hidden photon increases the binding energy of the $s$-state compared to the $p$-state.
This makes the Lamb-shift in muonic hydrogen more negative, in line with the observed effect.

Encouraged by this 
we would like to do a fit using the Lamb shift in ordinary and muonic hydrogen to fit the proton radius and the kinetic mixing parameter $\chi^2$ of the hidden photon. We can use the same strategy as outlined in Sect.~\ref{Constraints Using Atomic Spectra}, just including the proton radius as an additional parameter\footnote{As we are dealing with Lamb shifts (i.e transitions between states of the same $n$), changes in $\alpha$ are a subdominant effect (see Sect.~\ref{Naive Bounds for chi}).}.
The hidden photon contribution to the Lamb shift is already given in Eq.~\eqref{lambshift}. For muonic hydrogen we just have to replace
the electron mass with the muon mass. The lowest order dependence of the Lamb shift on the proton radius is given in Eq.~\eqref{rpdep}.

Unfortunately, the required values for $\chi^2$ are smaller than zero and since $\chi$ is a real parameter this rules out a hidden photon explanation. Why is this the case?
From Eqs.~\eqref{rpdep} and \eqref{lambshift} we see that from the perspective of the ($n=2$) Lamb shifts a shift in the proton radius by 
$\Delta r^{2}_{p}$ is equivalent to a non-vanishing $\chi^2$ for
\begin{equation}
\Delta r^{2}_{p}= -6\chi^2\frac{a^{4}_{o}m^{2}_{\gamma^{\prime}}}{(1+a_{o}m_{\gamma^{\prime}})^{4}}.
\end{equation}
where $a_{o}=1/(\alpha m_{o})$ is the Bohr radius of the orbiting particle.

This means that if we have non-vanishing $\chi^2>0$ we measure effectively measure a smaller proton radius in the Lamb shift measurement. 
Obviously both the ordinary and the muonic hydrogen measurements are affected in the same direction. Now, one can easily check that the effect is actually always bigger for larger $a_{o}$. In other words if there is a non-vanishing $\chi^2>0$ then the observed proton radius in
the Lamb shift of ordinary hydrogen should be even smaller than the one observed in muonic hydrogen. This is exactly the opposite of what is observed 
in Ref.~\cite{citeulike:7426442}.

Finally, we can use the same two measurements to form a bound, independent of electron scattering determinations
of the proton radius. However, in light of the fact that the two measurements are not consistent with each other we need to inflate our uncertainty similar to Eq.~\eqref{deltaH}.  Taking $M_{1}$ as the Lamb shift in atomic hydrogen and using Refs. \cite{PhysRevLett.82.4960, Simon:1980hu} we have\footnote{The $\delta M_{i}$ are calculated using the CODATA~\cite{mohr-2007} mean value for the proton radius of $r_p=0.8768$~fm.} $\delta M_{1} = 3 \times 10^{-11}$ eV and $\Delta M^{\star}_{1}=10^{-10}$~eV at the 1 $\sigma$ level\footnote{Note that since in this section we explicitly consider variations in $r_{p}$ in the bound, we could in principle ignore any contributions from variations in $r_{p}$ when calculating the theoretical uncertainty. However, as mentioned before, we will inflate our error to account for the inconsistency of the two measurements.}. Taking $M_{2}$ as the muonic hydrogen case, and using the experimental and theoretical values from \ref{Muonic Atoms} we get $\delta M_{2} = -3.11 \times 10^{-4}$ eV and $\Delta M^{\star}_{2}=6\times 10^{-4}$~eV. The corresponding bound is shown as the solid blue line in Fig.~\ref{specbounds}.

\section{Conclusions}
\label{Conclusion}
Atomic spectra can provide a powerful and clean probe of Coulomb's law at atomic length scales.
In this paper we have considered a variety of transitions in atomic hydrogen, hydrogenlike ions and exotic atoms, probing a wide range of different length scales. Currently, the best bounds are obtained from the measurements of the Lamb shift in ordinary hydrogen, a combination of the $1s_{1/2}-2s_{1/2}$ and the $2s_{1/2} - 8s_{1/2}$ transitions in atomic hydrogen, as well as the Lamb shift in hydrogen-like helium ions. These bounds are shown in 
Fig.~\ref{hpcurrent} and can be taken more generally as a constraint on deviations from Coulomb's law of the form Eq.~\eqref{Vcoulomb}. This significantly increases the range of pure Coulomb's law tests.

Future measurements of true muonium ($\mu^{+}\mu^{-}$) could significantly increase the tested area towards shorter length scales and smaller deviations. Similarly, a bound from the $2s_{1/2}^{F=1} - 2p_{3/2}^{F=2}$ transition in muonic hydrogen could penetrate new parameter space, provided that a suitable independent value of the proton radius is determined.
This gives additional motivation to seek further high precision determinations of the proton radius.

The deviation from Coulomb's law, Eq.~\eqref{Vcoulomb}, and the bounds we obtain can also be interpreted as a probe of massive hidden photons kinetically mixing with the ordinary photon. With currently available data we find no improvement over existing bounds. We note, however, that these bounds are especially
clean and model independent as they do not depend on the stability of the hidden photon or the presence/absence of certain decay channels.
Moreover, above mentioned future measurements also have good potential to probe new parameter space for hidden photons, giving them significant discovery potential for new physics.

Finally, we have briefly investigated a recent measurement of the $2s_{1/2}^{F=1} - 2p_{3/2}^{F=2}$ transition in muonic hydrogen which found a value inconsistent with theoretical calculations and previous measurements of the proton radius. 
We have considered the hidden photon as a mechanism to explain this discrepancy and found that the measurements of the Lamb
shift in ordinary hydrogen are in conflict with this hypothesis.

\section*{Acknowledgements}
The authors would like to thank A.~Lindner, J.~Redondo and A.~Ringwald for interesting discussions and useful comments.
Moreover, the authors are grateful to B.~Batell and M.~Pospelov for pointing out a crucial sign error in the discussion of the muonic 
hydrogen anomaly.

\appendix 
\section{Bounds on minicharged particles from atomic spectra}
\label{Appendix MCP}

\begin{figure}
\begin{center}
\includegraphics[width=100mm]{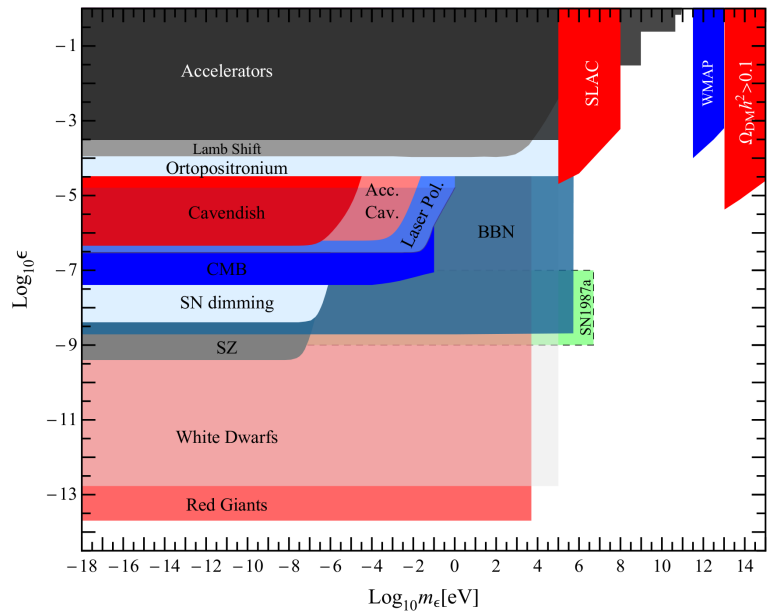}
\caption{Existing bounds for MCPs (see, e.g. \cite{Jaeckel:2010ni}). Note that the Lamb shift bound is produced using the method from \cite{Gluck:2007ia}. We modify this method to consider other atomic spectra. \label{MCPbounds}}
\end{center}
\end{figure}

Here we briefly examine bounds on minicharged particles (MCPs) produced by using atomic spectra. The existing bounds for MCPs are presented in Fig. \ref{MCPbounds}.

\begin{figure}
\begin{center}
\scalebox{0.4}[0.4]{
 \begin{picture}(130,282) (222,-127)
    \SetWidth{1.0}
    \SetColor{Black}
    \Photon(288,-70)(288,-102){7.5}{2}
    \SetWidth{3.0}
    \COval(288,-102)(22.627,22.627)(-135.0){Black}{White}\Line(299.314,-90.686)(276.686,-113.314)\Line(299.314,-113.314)(276.686,-90.686)
    \SetWidth{1.0}
    \Line[arrow,arrowpos=0.5,arrowlength=5,arrowwidth=2,arrowinset=0.2](240,154)(288,122)
    \Line[arrow,arrowpos=0.5,arrowlength=5,arrowwidth=2,arrowinset=0.2](288,122)(336,154)
    \Photon(288,-70)(288,-38){7.5}{2}
    \Photon(288,122)(288,90){7.5}{2}
    \Arc[arrow,arrowpos=0.5,arrowlength=5,arrowwidth=2,arrowinset=0.2](287,25)(63.977,335,695)
  \end{picture}
  }
  \end{center}
  \caption{Vacuum polarization diagram leading to a deviation from Coulomb's law in the presence of minicharged particles.}
  \label{vacpol}
\end{figure}
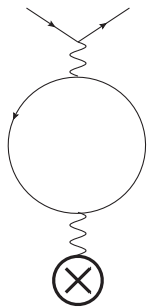
  
\begin{figure}
\begin{center}
\includegraphics[width=100mm]{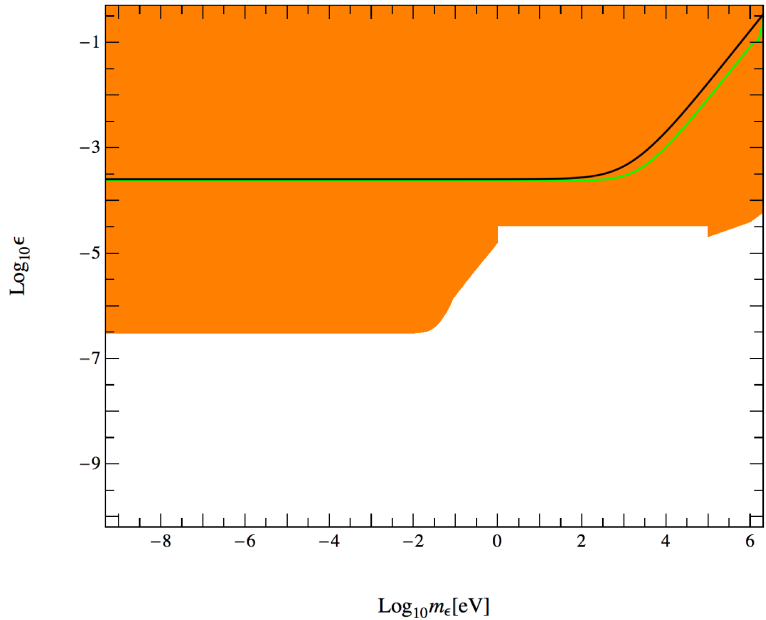}
\caption{The $1s_{1/2} - 2s_{1/2}$ (black) and $2s_{1/2} - 2p_{1/2}$~\cite{Gluck:2007ia} (green) bounds, both at the 2 $\sigma$ level. The $1s_{1/2} - 2s_{1/2}$ transition is renormalised using $2s_{1/2} - 8s_{1/2}$ as in the hidden photon case. For comparison we show previous bounds from purely laboratory experiments in orange. Our bounds do not penetrate new parameter space.
\label{twocoloursMCP}}
\end{center}
\end{figure}

As noted in~\cite{Jaeckel:2009dh} the vacuum polarization caused by minicharged particles (shown in Fig.~\ref{vacpol}) causes a modification to the Coulomb potential
\begin{equation}
V(\mathbf{r})=V_{\rm Coulomb}(r)+\delta V(r)_{MCP}
\end{equation}
where the additional term is the Uehling Potential, 
\begin{equation}
\label{deviationMCP}
\delta V(r)_{MCP} = 
\frac{Z\alpha}{r}\left[\frac{2\alpha\epsilon^2}{3\pi}\int^{\infty}_{2 m_{\epsilon} }dq\,\frac{\exp(-qr)}{q}\sqrt{1-\frac{4 m_{\epsilon}^2}{q^2}}\left(1+\frac{2 m_{\epsilon}^2}{q^2}\right)\right]
\end{equation}
Here $m_{\epsilon}$ is the mass of the minicharged particle, and $\epsilon$ the charge in units of $e$. In the limits of high and low mass $\delta$V(r) reduces as follows;
\begin{eqnarray}
\label{m,r}
\delta V(r)_{MCP} \approx \frac{Z\alpha}{r} \left[\frac{\alpha\epsilon^2}{4\sqrt{\pi}}\frac{\exp(-2 m_{\epsilon} r)}{(m_{\epsilon}r)^{\frac{3}{2}}}\right]
\quad\quad\quad\quad\quad\quad\quad\quad\quad\quad{\rm for}\quad  m_{\epsilon} \, r\gg 1,
\\\nonumber
\approx \frac{Z\alpha}{r}\left[-\frac{2\alpha\epsilon^2}{3\pi}\log(2  m_{\epsilon} r)-a\right],
\quad a\approx\frac{2\alpha\epsilon^2}{3\pi}\gamma \quad\quad\,\,\,\,{\rm for}\quad  m_{\epsilon} r\ll 1,
\end{eqnarray}
and where $\gamma$ is Euler's constant.

The potential decays exponentially in the high mass limit, as in the hidden photon case. However $\delta V(r)_{MCP}$ does not die off in the low mass limit. This makes sense physically, as a massless MCP will still be distinguishable from the photon and therefore will constitute a non-trivial modification to the photon propagator, which will result in observable effects. In fact we find that the strongest effects are observed in the low mass limit. This can be seen from saturation of the bounds in Fig.~\ref{MCPbounds}. Hence we obtain strong bounds for 
\begin{equation}
m_{\epsilon} \ll \frac{1}{l_{0}}
\end{equation}
where $l_{0}$ is the length scale of the experiment.

From \ref{m,r} we can see that the potential actually diverges logarithmically. We apply the same renormalisation procedure as we did with the hidden photon in Sect.~\ref{Renormalisation of alpha} to get
\begin{equation} \epsilon \leq \sqrt{\frac{\frac{n_{1} | \Delta M_{2} | } {M_{2}} + \frac{n_{2} | \Delta M_{1} | }{M_{1}}}{n_{1} f_{2}(m_{\epsilon}) - n_{2} f_{1}(m_{\epsilon})}}.
\end{equation}

Note the following limits;
\begin{itemize} 
\item{$m_{\epsilon} \rightarrow \infty$}: Here $f_{1}(m_{\epsilon}), f_{2}(m_{\epsilon}) \rightarrow$ 0 so $\epsilon$ increases and the bound dies off.
\item{$m_{\epsilon} \rightarrow 0$}: Here $f_{1}(m_{\epsilon})$ and $f_{2}(m_{\epsilon})$ both diverge logarithmically, but the divergences cancel and the denominator approaches a finite value as $m_{\epsilon} \rightarrow 0$, giving us a saturated bound. This shows that, as expected, effects of the MCP are maximised in the low mass limit.
\end{itemize}

We now apply this method to different atomic transitions. The analysis here is much briefer than in the hidden photon case, as we only consider transitions in atomic hydrogen. (The bounds from systems with higher $Z$ and exotic atoms are considerably weaker.)
As noted before, higher excited states (i.e states which do not involve $1s_{1/2}$ or $2s_{1/2}$) are difficult to measure and have high experimental uncertainties. Therefore we only consider transitions involving the $1s_{1/2}$ and $2s_{1/2}$ states. In Fig. \ref{twocoloursMCP} we plot the Lamb shift $2s_{1/2} - 2p_{1/2}$ \footnote{Note that this has already been considered in Ref.~\cite{Gluck:2007ia}}, and also a renormalised bound using the $1s_{1/2} - 2s_{1/2}$ and $2s_{1/2} - 8s_{1/2}$ transitions. The errors are the same as quoted in section \ref{Renormalised Method for chi}. We see that no improvement is found over existing bounds.

\section{Estimate for the theoretical uncertainty in the $2s_{1/2} - 2p_{1/2}$ transition in true muonium}
\label{Appendix Muonium Estimate}
The leading order contribution to the Lamb shift is proportional to the mass of the orbiting particle $m_{o}$. Therefore to get an approximate value of this transition we scale the atomic hydrogen value of by a factor $\sim \frac{m_{\mu}}{m_{e}}$ to get $\sim$ 1GHz.
\par We note that the major part of the uncertainty in the atomic hydrogen case comes from finite nuclear size effects, and that these are absent in true muonium. We can therefore get a naive estimate of the uncertainty in true muonium by subtracting the finite nuclear size contribution from the atomic hydrogen value \footnote{For values of the Lamb shift contributions in atomic hydrogen see, e.g. Ref.~\cite{Biraben}.} and scaling it up by $\frac{m_{\mu}}{m_{e}}$ to get $\sim$ 200 kHz.
\par However this naive estimate is inadequate. The reduced mass of the system is now larger, so that hadronic and muonic vacuum polarization contributions are now much more important. These effects must receive more careful treatment. To leading order~\cite{PhysRevA.59.4061}
\begin{equation}
E_{VP} \propto \left(\frac{m_{o}}{m_{e}}\right)^{2} \, m_{o} 
\end{equation}
so that we need to scale up vacuum polarization contributions in atomic hydrogen by a factor $(\frac{m_{\mu}}{m_{e}})^{3}$ which gives us a much larger contribution of $\sim$ 0.1 GHz. We therefore take this as our theoretical uncertainty, and set $\Delta M \sim$ 0.1 GHz.

\section{Bounds from hadronic atoms}
\label{Appendix Hadronic}
We review this option briefly, and argue that hadronic atoms currently do not produce strong bounds.

The existing candidates involve the $\pi^{-}$, $K^{-}$, $p^{+}$, and $K^{-}$ particles orbiting a proton nucleus.
In each case we find significant experimental uncertainties which immediately destroy bounds. For example with pionic hydrogen the experimental strong interaction shift of the $1s_{1/2}$ ground state \cite{Marton2007328c},
\begin{equation}
\epsilon_{1s} = \left(7.120 \pm 0.008 \pm 0.007 \right)\, \text{eV}
\end{equation}
where the errors are systematic and statistical respectively. This essentially means that, for example, the $3p_{1/2}-1s_{1/2}$ transition would have an error of at least $10^{-2}$ eV, which does not produce a useful bound.

All transitions in kaonic and sigmaonic hydrogen have large uncertainties due to the determination of the particle masses alone \cite{PDBook}.

Transitions in antiprotonic hydrogen have large uncertainties due to both strong interaction shifts and annihilation~\cite{antip}.

This is before we take into account theoretical uncertainties, for example finite size effects from both nucleus and orbiting particle, which we also expect to be large.

Overall we conclude that any benefits from the larger reduced masses of hadronic atoms are  washed out by experimental uncertainties and QCD effects.

\bibliographystyle{h-physrev3}
\bibliography{hiddenconstraints-final.bib}

\end{document}